\documentclass[12pt]{article}
\usepackage{epsf}
\usepackage{amsmath,amssymb}
\usepackage{graphicx}
\usepackage{color}
\usepackage{cite}
\usepackage{bm}
\usepackage[colorlinks=true,linkcolor=blue,citecolor=blue]{hyperref}
\usepackage{breakurl}
\usepackage{arydshln}
\usepackage{multirow}
\usepackage{aas_macros}
\usepackage{comment}
\usepackage{CJKutf8}
\usepackage{tikz}

\usepackage{braket}     
\usepackage{bm}

\renewcommand{\thefootnote}{\fnsymbol{footnote}}
\def\thefootnote{\fnsymbol{footnote}}

\makeatletter

\@addtoreset{equation}{section}
\makeatother

\newcommand{\ldef}{{} \overset{\mathrm{}}{{}:={}}{}}		
\newcommand{\ev}[1]{\Braket{#1} }   
\newcommand{\p}{{\bm{p}}}     
\newcommand{\q}{{\bm{q}}}     
\renewcommand{\r}{{\bm{r}}}
\renewcommand{\k}{{\bm{k}}}   
\newcommand{\x}{{\bm{x}}}     

\newcommand{\deltag}{\delta_{g}}
\newcommand{\tdeltag}{\tilde{\delta}_{g}}
\newcommand{\deltags}{\delta_{gs}}

\newcommand{\deltad}{\delta_{\rm D}}

\newcommand{\tdelta}{\tilde{\delta}}
\newcommand{\tdeltal}{\tilde{\delta}_{\rm L}}
\newcommand{\rI}{\textrm{I}}
\newcommand{\rII}{\textrm{I\hspace{-0.1em}I}}

\newcommand{\daa}{d_{\alpha\alpha}}
\newcommand{\daam}{d_{\alpha\alpha-}}
\renewcommand{\dag}{d_{\alpha\gamma}}
\newcommand{\dagm}{d_{\alpha\gamma-}}
\newcommand{\dgg}{d_{\gamma\gamma}}
\newcommand{\dxi}{d_{\xi}}
\newcommand{\dzeta}{d_{\zeta}}

\begin{document}

\begin{titlepage}

\flushright{KEK-TH-2635, KEK-Cosmo-0350}

\begin{center}

\vskip .75in

{\Large \bf Kurtosis consistency relation in large-scale structure as a probe of gravity theories }

\vskip .75in

{\large
Sora~Yamashita$\,^1$, 
Takahiko~Matsubara$\,^{2,3}$, 
Tomo~Takahashi$\,^4$, \\[4pt]
and 
Daisuke~Yamauchi$\,^5$,
}

\vskip 0.25in

{\em
$^{1}$Graduate School of Science and Engineering, Saga University, Saga 840-8502, Japan 
\vspace{2mm} \\
$^{2}$Institute of Particle and Nuclear Studies, High Energy Accelerator Research Organization (KEK), Oho 1-1, Tsukuba 305-0801, Japan 
\vspace{2mm} \\
$^{3}$The Graduate University for Advanced Studies (SOKENDAI), Tsukuba, Ibaraki 305-0801, Japan
\vspace{2mm} \\
$^{4}$Department of Physics, Saga University, Saga 840-8502, Japan
\vspace{2mm} \\
$^{5}$Department of Physics, Faculty of Science, Okayama University of Science, 1-1 Ridaicho, Okayama, 700-0005, Japan 
}

\end{center}
\vskip .5in

\begin{abstract}
Various gravity theories beyond general relativity have been
rigorously investigated in the literature such as Horndeski and
degenerate higher-order scalar-tensor (DHOST) theories. In general,
numerous model parameters are involved in such theories, which should
be constrained to test the theories with experiments and observations.
We construct the kurtosis consistency relations, calculated based on
matter density fluctuations, in which the information of gravity
theories is encoded. We derive two independent consistency relations
that should hold in the framework of the DHOST theories and argue
that such consistency relations would be useful 
for testing gravity
theories.

\end{abstract}

\end{titlepage}

\renewcommand{\thepage}{\arabic{page}}
\setcounter{page}{1}
\renewcommand{\thefootnote}{\#\arabic{footnote}}
\setcounter{footnote}{0}

\section{Introduction \label{sec:intro}}

Modification of gravity theories beyond general relativity (GR) has been rigorously studied, one of whose motivations is to explain the current accelerated cosmic expansion without resorting to a cosmological constant or dark energy. Modified gravity may also resolve some issues in cosmology such as the so-called Hubble tension and $S_8$ tension (see e.g., \cite{Khalife:2023qbu,Abdalla:2022yfr} 
for recent reviews). A variety of gravity theories have been developed and investigated (for recent reviews of modified gravity theories, see,
e.g., \cite{Arai:2022ilw,Shankaranarayanan:2022wbx}), 
and hence it would be imperative to test those theories with experiments and observations. In this paper, we consider the large-scale structure (LSS) of the Universe as such a cosmological probe, in particular, focusing on quasi-nonlinear growth of matter density fluctuations in modified gravity theories since such information in non-Gaussianity due to nonlinear growth of density fluctuations could capture characteristic signatures of modified gravity and may serve as a critical test.

Although various modified gravity theories have been investigated in the literature, the scalar-tensor family is the most intensively studied. In this paper, we focus on the degenerate higher-order scalar-tensor (DHOST) theories 
\cite{Langlois:2015cwa}, 
which include the Horndeski theory
\cite{Horndeski:1974wa,Deffayet:2011gz,Kobayashi:2011nu} 
and the Gleyzes-Langlois-Piazza-Vernizzi (GLPV) theory 
\cite{Gleyzes:2014dya,  Gleyzes:2014qga}. 
There have been many works on linear density perturbations in those theories, e.g.~
\cite{Langlois:2017mxy,Hiramatsu:2020fcd,Hirano:2019nkz},
furthermore, nonlinear properties of matter density fluctuations have also been analyzed: matter bi- and tri-spectra in the Horndeski theory were studied in
\cite{Takushima:2013foa,Takushima:2015iha}, 
and those in the GLPV and the DHOST theories were investigated in 
\cite{Hirano:2018uar,Hirano:2020dom}.

Although bi- and tri-spectra of matter density fluctuations are common observables to probe the non-Gaussian nature of density fluctuations, they are in general complicated functions of wave numbers, which may hinder the signature of modified gravity. Therefore in this paper we consider simpler observables, the skewness and kurtosis parameters as a measure of non-Gaussianity and discuss their predictions in modified gravity theories (see 
\cite{Matsubara:2003yt,Matsubara:2020knr} 
for the case of GR). Importantly, 
we can construct consistency relations\footnote{
Other types of consistency relations in LSS are also studied in \cite{Peloso:2013zw, Kehagias:2013yd, Creminelli:2013mca, Creminelli:2013poa, Valageas:2013cma, Rizzo:2016akm, Rizzo:2017zow, Crisostomi:2019vhj}.
} among the observables for the skewness and kurtosis parameters, which could give a useful test for modified gravity theories. Such relations can be derived by eliminating the bias parameters at higher orders which are usually considered to be difficult to measure. Indeed three of the present authors have derived consistency relations that hold among the skewness parameters in real and redshift space 
\cite{Yamauchi:2022fss}. 
In this paper, we extend the analysis done in 
\cite{Yamauchi:2022fss} 
to include the kurtosis parameters and then derive more consistency relations which would provide a further test for theories of gravity through LSS.

The structure of this paper is as follows. In Section~\ref{sec:matter_density}, 
we provide a summary of matter density fluctuation in the DHOST theory and introduce the local bias model. Then, in Section~\ref{sec:SkewAndKurt}, 
we define the skewness and kurtosis parameters adopted in this paper and derive the consistency relations by eliminating the bias parameters, which would give a critical test of gravity theories. In the final section, we conclude our paper and discuss the implications of our results for testing gravity theories.

\section{Matter density fluctuation in real space \label{sec:matter_density}}

In this section, we will first provide a summary of the formalism to discuss the evolution of matter density fluctuations in the DHOST theory. Then we also introduce the bias model adopted in this paper.

\subsection{Evolution of the matter density fluctuation\label{subsec:EvolutionOfMatter}} 

Matter density fluctuations $\delta(t, \x)$,  which depend on cosmic time and spatial coordinates, are defined by matter density $\rho(t,\x)$ and its spatial average $\bar\rho(t)$:
\begin{align}
\delta(t, \x)
\ldef \frac{\rho(t,\x)-\bar\rho(t)}{\bar\rho(t)}.
\end{align}
It would be convenient to consider it in the Fourier space as
\begin{align}
\tdelta(t,\k) 
\ldef \int d^3x\, e^{-i\k\cdot\x} \,
\delta(t, \x)  \,.
\end{align}
The evolution of the matter fluctuations $\tdelta(t, \k)$ can be obtained by solving fluid equations perturbatively in the Friedmann-Robertson-Walker (FRW) Universe.  For brevity, below we omit $t$ in the arguments of $\delta(t, \x)$ and $\tdelta(t, \k)$. The Fourier transform of the matter density fluctuations can be formally expanded in terms of the linear matter density fluctuations as
\begin{align}
	\tdelta(\k)
	={}&
		\sum_{n=1}^{\infty}
			\int \frac{d^3k_1}{(2\pi)^3}\cdots 
            \frac{d^3k_n}{(2\pi)^3}
			(2\pi)^3\deltad(\k_{12\dots n}-\k)
            \frac{F_n(\k_1,\ldots,\k_n)}{n!}
			\tdeltal(\k_1)\cdots\tdeltal(\k_n) ,
		\label{delta-fourier}
\end{align}
where $\tdeltal(\k)$ is a linear solution of the fluid equations and $\k_{12\dots n}=\k_1 + \k_2 +\dots+\k_n$.  The modification of gravity theory alters the clustering property of the nonlinear structure. Hence, the time- and wavenumber-dependence of the kernel functions $F_n$ yields a powerful probe of modified gravity theories.  In this paper, we only focus on the DHOST theories, while there is a wide variety of gravity theories that yield different signatures to the nonlinear kernels.  In the case of the DHOST theories, the kernels $F_n$ up to the third order can be written
as~
\cite{Takushima:2013foa,Takushima:2015iha,Hirano:2020dom}
\begin{align}
    &F_1 = 1 \,, \\
	&\frac{F_2(t,\k_1,\k_2)}{2!} = \kappa(t) \alpha_s(\k_1,\k_2)
		-\frac{2}{7}\lambda(t) \gamma(\k_1,\k_2)\,,\label{eq:F2} \\
	&
        \begin{aligned}
        \frac{F_3(t,\k_1,\k_2,\k_3)}{3!}
		={} &{d_{\alpha\alpha}(t)}\alpha\alpha(\k_1,\k_2,\k_3) 
		-\frac{4}{7}{d_{\alpha\gamma}(t)}\alpha\gamma(\k_1,\k_2,\k_3) 
		-\frac{2}{21}{d_{\gamma\gamma}(t)}\gamma\gamma(\k_1,\k_2,\k_3) 
            \\
		+&{d_{\alpha\alpha-}(t)}\alpha\alpha_-(\k_1,\k_2,\k_3) 
		+{d_{\alpha\gamma-}(t)}\alpha\gamma_-(\k_1,\k_2,\k_3) 
            +\frac{1}{9}{d_{\xi}(t)}\xi_c(\k_1,\k_2,\k_3) \\
            +&
            {d_\zeta(t)}\zeta_c(\k_1,\k_2,\k_3) \,,\label{eq:F3}
        \end{aligned}
\end{align}
where the functions $\alpha_s, \gamma, \alpha\alpha,\ldots$ represent the mode-coupling and their explicit forms are defined in the following equations:
\begin{align}
&\begin{aligned}
    &\alpha(\k_1,\k_2)\ldef 1+\frac{\k_1\cdot\k_2}{k_1^2},\\
    &\gamma(\k_1,\k_2) \ldef 1 - (\hat\k_1\cdot\hat\k_2)^2 ,\\
    &\alpha_s(\k_1,\k_2) \ldef \frac12[\alpha(\k_1,\k_2) + \alpha(\k_2,\k_1)],\\
    &\alpha_{as}(\k_1,\k_2) \ldef \frac12[\alpha(\k_1,\k_2) - \alpha(\k_2,\k_1)],\\
\end{aligned}&
&\begin{aligned}
    &\alpha\alpha(\k_1,\k_2,\k_3) \ldef \alpha_s(\k_1,\k_2+\k_3)\alpha_s(\k_2,\k_3),\\
    &\alpha\gamma(\k_1,\k_2,\k_3) \ldef \alpha_s(\k_1,\k_2+\k_3)\gamma(\k_2,\k_3) ,\\
    &\gamma\alpha(\k_1,\k_2,\k_3) \ldef \gamma(\k_1,\k_2+\k_3)\alpha_s(\k_2,\k_3) ,\\
    &\gamma\gamma(\k_1,\k_2,\k_3) \ldef \gamma(\k_1,\k_2+\k_3)\gamma(\k_2,\k_3) ,\\
    &\alpha\alpha_-(\k_1,\k_2,\k_3) \ldef \alpha_{as}(\k_1,\k_2+\k_3)\alpha_s(\k_2,\k_3) ,\\
    &\alpha\gamma_-(\k_1,\k_2,\k_3) \ldef \alpha_{as}(\k_1,\k_2+\k_3)\gamma(\k_2,\k_3),
\end{aligned}\label{eq:functions1}
\\
&\begin{aligned}
    &\xi(\k_1,\k_2,\k_3) \ldef 
        1-3(\hat\k_1\cdot\hat\k_2)^2 +2(\hat\k_1\cdot\hat\k_2)(\hat\k_2\cdot\hat\k_3)(\hat\k_3\cdot\hat\k_1), \\
    &\xi_c(\k_1,\k_2,\k_3) \ldef
        \frac13\left[\xi(\k_1,\k_2,\k_3)+\xi(\k_2,\k_3,\k_1)+\xi(\k_3,\k_1,\k_2)\right] ,\\
    &\zeta(\k_1,\k_2,\k_3) \ldef
	(\hat\k_1\cdot\hat\k_2)\left[(\hat\k_1\cdot\hat\k_2)+\frac{k_2}{k_1}\right]+
        \left(2\frac{k_2}{k_3}+\frac{k_2^2+k_3^2}{k_3k_1}\right)
	(\hat\k_1\cdot\hat\k_2)(\hat\k_2\cdot\hat\k_3),\\
    &\zeta_c(\k_1,\k_2,\k_3) \ldef\frac13\left[\zeta(\k_1,\k_2,\k_3)+\zeta(\k_2,\k_3,\k_1)+\zeta(\k_3,\k_1,\k_2)\right].
\end{aligned}\hspace{-50em}
\label{eq:functions2}
\end{align}
These functions are not independent in the sense of the convolution, as we will show in the footnote~\ref{footnote:3}.
In particular, we have used $\xi_c$ instead of $\gamma\alpha$ to describe Eq.~\eqref{eq:F3}. The time-dependent parameters, $\kappa$ and $\lambda$ for the second-order kernel, 
$d_{\alpha\alpha}$, $d_{\alpha\alpha -}$,
$d_{\alpha\gamma}$, $d_{\alpha\gamma -}$, 
$d_{\gamma\gamma}$, $d_{\xi}$, and $d_{\zeta}$  
for the third-order kernel, which appears in Eqs.~\eqref{eq:F2} and \eqref{eq:F3}, are model parameters that depend on the underlying gravity theory. In the limit of the Einstein-de Sitter (EdS) Universe in general relativity, one finds 
\begin{align}
\kappa 
= \lambda
=d_{\alpha\alpha}
=d_{\alpha\gamma}
=d_{\gamma\gamma}
= d_\xi = 1  \,,
\qquad
 d_{\alpha\alpha -}
=d_{\alpha\gamma -}
= d_\zeta 
=0 
\qquad
{\rm (EdS)}
\,.
\end{align}
In the case of the Horndeski theory, we still have 
\begin{align}
 \kappa=d_{\alpha\alpha}=1 \,,
 \qquad
 d_{\alpha\alpha-}=d_{\alpha\gamma-}=d_{\zeta}=0 
 \qquad
 {\rm (Horndeski)}\,,
\end{align}
but $\lambda$, $d_{\alpha\gamma}$, $d_{\gamma\gamma}$, and $d_\xi$ can deviate from unity~
\cite{Takushima:2013foa,Takushima:2015iha,Yamauchi:2017ibz}. When the gravity is described by the DHOST theory (beyond Horndeski),
all the nine parameters can deviate from those in the EdS or Horndeski case~\cite{Hirano:2020dom}. In particular, if observations show the deviation of either $\kappa$ or $d_{\alpha\alpha}$ from unity, and/or $d_{\alpha\alpha-}, d_{\alpha\gamma-}$ or $d_\zeta$ from zero, the Horndeski family would be ruled out.

\subsection{Bias model \label{subsec:bias}}

In section \ref{subsec:EvolutionOfMatter}, we have summarized the quasi-nonlinear solution of $\delta(\x)$ that is dependent on gravity theories. However, it is important to note that evolutions of structures in the Universe are governed not only by the gravitational interaction but also by other interactions involving baryonic components. These interactions have a significant impact on the formation of stars and galaxies and are the main reasons why galaxies are biased objects of matter distribution. Unfortunately, incorporating these effects can be quite challenging.  
However, in order to include the bias effects, we can customarily use bias models that describe the galaxy distribution $\deltag(\x)$ as a functional $\mathcal F$ of $\delta(\x)$:
\begin{align}
    \deltag(\x) = \mathcal F[\delta(\x)].
\end{align}
In this paper, we consider the local bias model\footnote{
We assumed tidal bias can be negligible for simple discussion. 
Inclusion of the tidal term would be straightforward.
} \cite{Desjacques:2016bnm, McDonald:2009dh} 
which assumes the following form for $\cal F$:
\begin{align}
    \deltag(\x) =
    &
    b_1\left[
    \delta(\x)
    +\frac{1}{2!}\beta_2 \delta^2(\x) 
    + \beta_{K^2}\sum_{i,j} K_{ij} K_{ji}
    \right.
    \nonumber \\
    & \left.\,
    +\frac{\beta_3}{3!}\delta^3(\x) + \beta_{K^3}\sum_{i,j,k}
		K_{ij}K_{jk}K_{ki}
	+\beta_{\delta K^2}\delta(\x)\sum_{i,j}
	K_{ij}K_{ji}
	+\mathcal O(\delta^4) 
    \right].
	\label{delta-bias-real}
\end{align}
where 
$\{b_1,$ $\beta_2,$ $\beta_{K^2}\}$ are the bias parameters for the second order, and $\{\beta_3,$ $\beta_{K^3},$ $\beta_{\delta K^2}\}$ are those for the third order and
$K_{ij} \ldef \left[\partial_i\partial_j/\partial^2 
- \delta_{ij}/3\right]\delta(\x)$.

Using Eq.~\eqref{delta-fourier}, the Fourier transform of \eqref{delta-bias-real} can be written as 
\begin{align}
	\tdeltag(\k) = {}&{} b_1\tdeltal(\k) \nonumber \\
	{}&{}+ b_1\int\frac{d^3k_1}{(2\pi)^3}
        \frac{d^3k_2}{(2\pi)^3}
        (2\pi)^3\deltad(\k_{12}-\k)
        \frac{Z_2(\k_1,\k_2)}{2!b_1}
        \tdeltal(\k_1)\tdeltal(\k_2)
		\nonumber \\
	{}&{}+b_1\int\frac{d^3k_1}{(2\pi)^3}
        \frac{d^3k_2}{(2\pi)^3}\frac{d^3k_3}{(2\pi)^3}
        (2\pi)^3\deltad(\k_{123}-\k)
        \frac{Z_3(\k_1,\k_2,\k_3)}{3!b_1}
        \tdeltal(\k_1)\tdeltal(\k_2)\tdeltal(\k_3) \nonumber \\
    &+ \mathcal O(\tdeltal^4),
  \label{tdeltag-dhost}
\end{align}
where
\begin{align}
    &{}\frac{Z_2(\k_1,\k_2)}{2!b_1} = 
	\frac{F_2(\k_1,\k_2)}{2!} + \frac{\beta_2}{2!} 
	+ \beta_{K^2}\left[
		(\hat\k_1\cdot\hat\k_2)^2 - \frac13
	\right]\,,\label{eq:Z2} \\
    &{}\frac{Z_3(\k_1,\k_2,\k_3)}{3!b_1} = 
	\frac{F_3(\k_1,\k_2,\k_3)}{3!} 
        +\frac{\beta_3}{3!}
        +2\beta_{K^2}\frac{F_2(\k_2,\k_3)}{2!}\left[ 
            \left( \hat\k_1\cdot \hat\k_{23} \right)^2-\frac13 
        \right]
	+ \beta_2\frac{F_2(\k_1,\k_2)}{2!}     \nonumber \\
    {}&{} \qquad\qquad\qquad
	+\beta_{K^3}\left[
		(\hat\k_1\cdot\hat\k_2)(\hat\k_2\cdot\hat\k_3)(\hat\k_3\cdot\hat\k_1)
		-(\hat\k_1\cdot\hat\k_2)^2+\frac29
	\right]
        +\beta_{\delta K^2}\left[
            (\hat\k_1\cdot\hat\k_2)^2-\frac13
        \right].\label{eq:Z3}
\end{align}
We note that thanks to the symmetry of the integrands, the kernels can be written in the asymmetric functional form for the permutation of their arguments $\{ \bm k_i \}$, which we utilize in our calculations in the next section.

We then introduce a Gaussian window function as a smoothing function as
\begin{align}
W_R(x) \ldef 
\frac{1}{(2\pi R^2)^{3/2}}
\exp[-x^2/(2R^2)] \,,
\label{eq:Gaussian smoothing}
\end{align}
and the density fluctuations are smoothed as:
\begin{align}
    \deltag(\x)
    \,\longrightarrow\,
    \deltags(\x ;R)= \int d^3x' \deltag(\x')W_R(|\x-\x'|)
    \,,\label{eq:delta_gs}
\end{align}
where $R$ is the so-called smoothing scale.  The Fourier transform of the smoothed density fluctuation is given by $\tilde\delta_{gs} ({\bm
  k})=\tilde\deltag ({\bm k})W(kR)$, where $W(kR)$ denotes 
  the Fourier transformed
  smoothing window function.  In the case of the
Gaussian smoothing given in Eq.~\eqref{eq:Gaussian smoothing}, we have
$W(kR)=e^{-k^2R^2/2}$,
and thus we obtain
\begin{align}
    \deltags(\x ;R)
    =\int \frac{d^3k}{(2\pi)^3}\,
    e^{i\k\cdot\x}\tdeltag(\k)e^{-k^2R^2/2} \,,
\end{align}
The perturbation theory 
gives a good description on scales with $k^{-1} \gtrsim 20\,\mathrm{Mpc}/h$ \cite{Matsubara:2020knr}, where $h$ is the dimensionless Hubble constant defined by $h$$\ldef$$H_0/(100~\mathrm{km/s/Mpc})$.

Taking the smoothing and the bias effects into account, 
we define the variance and the spectral moment as 
\begin{align}
    \sigma_0^2(R) \ldef \ev{\deltags^2(\x;R)}
    \,,\ \ \ 
     \sigma_1^2(R) \ldef \ev{[\nabla\deltags(\x;R)]^2}
     \,.\label{eq:sigma_0 sigma_1}
\end{align}
In the lowest-order approximation of the perturbation theory, these can be reduced to
\begin{align}
    &\begin{aligned}[b]
        \sigma_0^2(R) 
          &=\frac{b_1^2}{2\pi^2}\int dk\, k^2 P_{\rm L}(k)e^{-k^2R^2}, 
    \end{aligned}\label{sigma0}\\
    &\begin{aligned}[b]
        \sigma_1^2(R) 
                    &=\frac{b_1^2}{2\pi^2}\int dk\, k^4 P_{\rm L}(k)e^{-k^2R^2}.
    \end{aligned}
    \label{sigma1}
\end{align}
Here, $P_{\rm L}(k)$ represents the power spectrum of the linear matter density fluctuation, which is defined with correlation function:
\begin{align}
        \ev{\tdeltal(\k)\tdeltal(\k')}_c
        =(2\pi)^3\deltad (\k+\k')P_{\rm L}(k),
\end{align}
where $\ev{\cdots}_c$ indicates a two-point cumulant defined with ensemble average.

\section{Skewness and kurtosis consistency relation \label{sec:SkewAndKurt}}

By using the observed density fluctuations of matter introduced in the previous subsection, we can calculate the skewness and kurtosis parameters for the galaxy number density fluctuations. In this section, we first define the skewness and kurtosis parameters and give a general formula in the DHOST theory. Then we derive the consistency relations for the model parameters, which can serve as a test of gravity theories.

\subsection{Skewness and kurtosis parameters
\label{sec:skewness_kurtosis_parameters}}
Although galaxy distribution is a crucial sample for studying gravity, it includes unknown bias parameters, and we cannot test gravity theories easily. 
For studying theories of gravity, we should derive a quantity or relation that does not involve unknown bias parameters. 
To do this, we first define three skewness and five kurtosis parameters using the smoothed galaxy number density fluctuation $\delta_{gs}$ including the bias parameters.  
Then we derive the eight equations for these skewness and kurtosis parameters by using the model and bias parameters, from which we can express the bias parameters as functions of skewness and kurtosis parameters.
As we will see later, since we only need two skewness and
three kurtosis equations to remove the unknown bias parameters, three conditions (one skewness and two kurtosis relations) remain, which can be regarded as the consistency relations among the model parameters with the linear bias $b_1$. The purpose of our presentation is to show the methodology of constraining gravity theories from skewness and kurtosis parameters by removing bias uncertainty in a simple model of bias. It is possible to consider more complicated bias models and to increase the number of bias parameters, as one can increase the number of independent skewness and kurtosis parameters with higher-order derivatives.

With the use of the smoothed galaxy number density fluctuation
$\delta_{gs}$, Eq.~\eqref{eq:delta_gs}, the variance $\sigma_0$ and the spectral moment $\sigma_1$,  Eq.~\eqref{eq:sigma_0 sigma_1}, the
skewness and kurtosis parameters in real space can be defined by introducing some
spatial derivatives as~\cite{Matsubara:2020knr}
\begin{flalign}
&\begin{aligned}
	&S^{(0)}_g \ldef \frac{\ev{\delta_{gs}^3}_c}{\sigma_{0}^4}, \\
	&S^{(1)}_g \ldef \frac32 \cdot 
    \frac{\ev{\delta_{gs}|\nabla\delta_{gs}|^2}_c}
    {\sigma_{0}^2\sigma_{1}^2},
	\\
	&S^{(2)}_g \ldef -\frac94\cdot 
    \frac{\ev{|\nabla\delta_{gs}|^2\Delta\delta_{gs}}_c}{\sigma_{1}^4},
\end{aligned}& 
&\begin{aligned}
	&K^{(0)}_g\ldef \frac{\ev{\delta_{gs}^4}_c}{\sigma_{0}^6} \,,\\
	&K^{(1)}_g\ldef 2\cdot 
        \frac{\ev{\delta_{gs}^2|\nabla\delta_{gs}|^2}_c}
        {\sigma_{0}^4\sigma_{1}^2} \,,\\
	&K^{(2_\rI)}_{g}\ldef -\frac35 \cdot 
        \frac{5\ev{\delta_{gs}|\nabla\delta_{gs}|^2\Delta\delta_{gs}}_c+
		\ev{|\nabla\delta_{gs}|^4}_c}{\sigma_{0}^2\sigma_{1}^4} \,,\quad\\
	&K^{(2_\rII)}_{g}\ldef -\frac35 \cdot 
        \frac{5\ev{\delta_{gs}|\nabla\delta_{gs}|^2\Delta\delta_{gs}}_c+
		3\ev{|\nabla\delta_{gs}|^4}_c}{\sigma_{0}^2\sigma_{1}^4},\\
	&K^{(3)}_g\ldef 9\cdot
        \frac{\ev{|\nabla\delta_{gs}|^2(\Delta\delta_{gs})^2}_c
		-\ev{|\nabla\delta_{gs}|^2\delta_{gs,ij}\delta_{gs,ij}}_c}{\sigma_1^6} \,,
  \label{skewness-kurtosis-def}
\end{aligned}&&
\end{flalign}
where 
$\delta_{gs,ij} \ldef \partial_i\partial_j\delta_{gs}$.  
Using the perturbative expression of the smoothed galaxy number density fluctuations Eq.~\eqref{tdeltag-dhost} and the lowest order approximation, we can recast the skewness parameters defined in Eq.~\eqref{skewness-kurtosis-def} as the following form:
\begin{flalign}
    S^{(a)}_g &= \frac{6{b_1^3}}{\sigma_0^{4-2a}\sigma_1^{2a}}
    \int\frac{d^3k_1}{(2\pi)^3}\frac{d^3k_2}{(2\pi)^3}
    \frac{d^3k_3}{(2\pi)^3}(2\pi)^3\deltad(\k_{123})
    s^{(a)}(\k_1,\k_2,\k_3)
    e^{-(k_1^2+k_2^2+k_3^2)R^2/2}
    &&\nonumber \\
    &\qquad\qquad\qquad\times
    \frac{Z_2(\k_1,\k_2)}{2!b_1}P_{\rm L}(k_1)P_{\rm L}(k_2) \,,
    \qquad(a=0,1,2)
    \label{eq:S_g}
\end{flalign}
where $s^{(a)}(\k_1,\k_2,\k_3=-\k_{12})$ is defined as
\begin{align}
    &s^{(0)} = 1 ,\\
    &s^{(1)} = \frac12(k_1^2+k_2^2+\k_1\cdot\k_2),\\
    &s^{(2)} = \frac32\left[k_1^2k_2^2 - (\k_1\cdot\k_2)^2\right].
\end{align}
To perform the integration in Eq.~\eqref{eq:S_g},
we have integrated for $\k_3$ with Dirac's delta function and then transformed valuables:
\begin{align}
    &\k_1\rightarrow \p/R=(p\sin\theta, 0, p\cos\theta)/R, \\
    &\k_2\rightarrow \q/R=(0,0,q)/R \,,
\end{align}
then Eq.~\eqref{eq:S_g} is rewritten as
\begin{flalign}
   S^{(a)}_g 
    &=
    \frac{6{b_1^3}}{8\pi^4R^{2a+6}\sigma_0^{4-2a}\sigma_1^{2a}}
    \int_0^\infty p^2q^2 dpdq ~
    \mathcal S^{(a)}(p,q)
    e^{-p^2-q^2}
    P_{\rm L}(p/R)P_{\rm L}(q/R),
    \qquad(a=0,1,2)
    \label{skewness-specific}
\end{flalign}
where $\mathcal S^{(a)}$ is defined as
\begin{align}
    &\mathcal S^{(a)}=
    \int_0^\pi \sin\theta d\theta~
    s^{(a)}(\p,\q,-\p-\q)
    \frac{Z_2(\p,\q)}{2!b_1} 
    e^{-pq\cos\theta} \,.
\end{align}
When we rewrote the formula for the skewness parameters,
we adapted the following relations:
\begin{align}
    &\frac{Z_2(\p/R,\q/R)}{2!b_1} = \frac{Z_2(\p,\q)}{2!b_1}, \\
    &s^{(a)}(\p/R,\q/R,-\p/R-\q/R)
    =R^{-2a}s^{(a)}(\p,\q,-\p-\q).
\end{align}

Similarly, we can rewrite the kurtosis parameters defined in Eq.~\eqref{skewness-kurtosis-def} as
\begin{flalign}
    K^{(a)}_g&=\frac{24{b_1^4}}{\sigma_0^{6-2a}\sigma_1^{2a}}
    \int\frac{d^3k_1}{(2\pi)^3}\frac{d^3k_2}{(2\pi)^3}
    \frac{d^3k_3}{(2\pi)^3}\frac{d^3k_4}{(2\pi)^3}
    (2\pi)^3\deltad(\k_{1234}) 
    \nonumber
    \\
    &\qquad\times
    \kappa^{(a)}(\k_1,\k_2,\k_3,\k_4)
    e^{-(k_1^2+k_2^2+k_3^2+k_4^2)R^2/2}
    \nonumber \\
    &\qquad\times
    \left[
        \frac{Z_3(\k_1,\k_2,\k_3)}{3!b_1}
         P_{\rm L}(k_3)
    +
	2\frac{Z_2(\k_1,\k_{23})}{2!b_1}
        \frac{Z_2(\k_2,-\k_{23})}{2!b_1}
        P_{\rm L}(k_{23})
	\right]P_{\rm L}(k_1)P_{\rm L}(k_2)
 \nonumber \\
 &
 \qquad
 (a=0,1,2_{\rI},2_{\rII},3)
 \label{eq:K_g} \,, 
\end{flalign}
where $\kappa^{(a)}(\k_1,\k_2,\k_3, \k_4 =-\k_{123})$ is defined as
\begin{align}
    &\kappa^{(0)} = 1 ,\\
    &\kappa^{(1)} = \frac13(k_1^2+k_2^2+k_3^2+
        \k_1\cdot\k_2+\k_2\cdot\k_3+\k_3\cdot\k_1),\\
    &\kappa^{(2_{\rI})} = 
        \frac{1}{10}\left\{
        5\left[k_1^2k_2^2-(\k_1\cdot\k_2)^2\right]
        -6(\k_1\cdot\k_3)(\k_2\cdot\k_3)+2(\k_1\cdot\k_2)k_3^2
        \right\} + \textrm{cyc.},\\
    &\kappa^{(2_{\rII})} =
    \frac{1}{10}\left\{
        5\left[k_1^2k_2^2-(\k_1\cdot\k_2)^2\right]
        +2(\k_1\cdot\k_3)(\k_2\cdot\k_3)+6(\k_1\cdot\k_2)k_3^2
        \right\} + \textrm{cyc.},\\
    &\kappa^{(3)} =
    \frac32\left[
    k_1^2k_2^2k_3^2+2(\k_1\cdot\k_2)(\k_2\cdot\k_3)(\k_3\cdot\k_1)
    -3(\k_1\cdot\k_2)^2k_3^2\right]+ \textrm{cyc.} \,.
    \label{kappa^3 def}
\end{align}
Here ``$+\textrm{cyc.}$'' 
indicates the terms with $\bm k$'s being permutated.  We note that all of the above $\kappa^{(a)}$ are completely symmetric for any permutation of their arguments. 
The integration of Eq.~\eqref{eq:K_g} can be performed by introducing the new variables as
\begin{align}
    &\k_1\rightarrow \p/R=(p\sin\theta\cos\phi, p\sin\theta\sin\phi, p\cos\theta)/R, \\
    &\k_2\rightarrow \q/R=(q\sin\theta',0,q\cos\theta')/R, \\
    &\k_3\rightarrow (\r-\q)/R = (0,0,r)/R-(q\sin\theta',0,q\cos\theta')/R \,,
\end{align}
and integrate $\k_4$ with Dirac's delta function, from which we obtain
\begin{flalign}
    K^{(a)}_g &=
    \frac{24{b_1^4}}{32\pi^6R^{2a+9}\sigma_0^{6-2a}\sigma_1^{2a}}
    \int_0^\infty p^2q^2r^2dpdqdr~e^{-p^2-q^2-r^2} \nonumber\\
    &\qquad\times\Biggl[
        \int_0^{2\pi} \frac{|\sin\theta'|}{2}d\theta'~
        \mathcal L_{1}^{(a)}(p,q,r,\theta')e^{qr\cos\theta'}
        P_{\rm L}(\sqrt{q^2+r^2-2qr\cos\theta'}/R)
    \nonumber \\
    &\qquad\quad
    +2\mathcal L_{2}^{(a)}(p,q,r)
    P_{\rm L}(r/R)
    \Biggr]P_{\rm L}(p/R)P_{\rm L}(q/R),
    \qquad(a=0,1,2_{\rI},2_{\rII},3)
\label{kurtosis-specific}
\end{flalign}
where $\mathcal L_1^{(a)}$ and $\mathcal L_2^{(a)}$ are defined as
\begin{align}
    &\mathcal L_1^{(a)}(p,q,r,\theta') =  
    \int_0^\pi \sin\theta d\theta~
    \kappa^{(a)}(\p,\q,\r-\q,-\p-\r)
    \frac{Z_3(\p,\q,\r-\q)}{3!b_1} \, e^{-pr \cos \theta} ,
    \\
    &\begin{aligned}[b]
        \mathcal L_2^{(a)}(p,q,r) =
        \int_0^\pi \sin\theta d\theta\int_0^{2\pi}\frac{|\sin\theta'|}{2} d\theta'~&
        \kappa^{(a)}(\p,\q,\r-\q,-\p-\r)
        \\
        &\quad\times\frac{Z_2(\p,\r)}{2!b_1}\frac{Z_2(\q,-\r)}{2!b_1}
    e^{qr\cos\theta'} e^{-pr \cos \theta},
    \end{aligned}
\end{align}
with applying the following relations:
\begin{align}
    &\frac{Z_3(\p/R,\q/R,\r/R-\q/R)}{3!b_1} = \frac{Z_3(\p,\q,\r-\q)}{3!b_1}, \\
    &\kappa^{(a)}(\p/R,\q/R,\r/R-\q/R,-\p/R-\r/R)
    =R^{-2a}\kappa^{(a)}(\p,\q,\r-\q,-\p-\r).
\end{align}

To proceed with the analysis, we need the explicit form of the kernels $Z_2$ and $Z_3$. When we substitute Eq.~\eqref{eq:F2} into Eq.~\eqref{eq:Z2}, the explicit form of $Z_2$ and the combination of $Z_2$ appearing in $Z_3$ (see Eq.~\eqref{eq:Z3}) can be written as follows:
\begin{align}
    &\frac{Z_2(\k_1,\k_2)}{2!b_1} 
    =
    h_0 + h_1(\k_1\cdot\k_2)\left(\frac{1}{k_1^2}+\frac{1}{k_2^2}\right)
    + h_2\frac{(\k_1\cdot\k_2)^2}{k_1^2k_2^2} ~,\\
    &\frac{Z_2(\k_1,\k_{23})}{2!b_1}\frac{Z_2(\k_2,-\k_{23})}{2!b_1} 
    \nonumber \\
    &\qquad 
    =
    \frac{1}{(k_1k_2)^2 |\k_{23}|^4}\nonumber \\
    &\qquad \quad\times
    \left\{
    |\k_{23}|^2 
    \left[-h_1(\k_2\cdot\k_3)+(h_0-h_1){k_2}^2\right]
    -(\k_2\cdot\k_{23}) 
    \left[-h_2(\k_2\cdot\k_3)+(h_1-h_2)k_2^2\right]
    \right\}\nonumber \\
    &\qquad \quad
    \times\left\{
    |\k_{23}|^2 \left[
    h_1(\k_1\cdot\k_{23})+h_0k_1^2\right]
    +(\k_1\cdot\k_{23})\left[
    h_2(\k_1\cdot\k_{23})+h_1 k_1^2\right]
    \right\} \,,
\end{align}
where 
\begin{align}
\begin{aligned}
&h_0=\frac{\beta_2}{2}-\frac{\beta_{K^2}}{3} + \kappa -\frac27\lambda, \\
&h_1=\frac\kappa2, \\
&h_2=\beta_{K^2} + \frac27\lambda.
\end{aligned}
\end{align}
Substituting Eq.~\eqref{eq:F3} into Eq.~\eqref{eq:Z3}, we obtain the 
explicit form of $Z_3$ as\footnote{\label{footnote:3}
We used the following relation:
\begin{align}
    &\int d^3k_1d^3k_2d^3k_3 \nonumber\\
    &\qquad\left[
    \alpha(\k_1,\k_{23})\gamma(\k_2,\k_3)
    -2\gamma(\k_1,\k_{23})\alpha(\k_2,\k_3)
    +2\gamma(\k_1,\k_{23})\gamma(\k_2,\k_3)
    \right]f(\k_1,\k_2,\k_3)
    \nonumber \\
    &\qquad\qquad=
    \int d^3k_1d^3k_2d^3k_3~
    \xi(\k_1,\k_2,\k_3)f(\k_1,\k_2,\k_3),\nonumber
\end{align}
where $f(\k_1,\k_2,\k_3)$ is some function 
with the symmetry for any permutation 
of $\k_1,~\k_2$ and $\k_3$. 
}
:
\begin{align}
    &\hspace{-3em}
    \frac{Z_3(\k_1,\k_2,\k_3)}{3!b_1} \nonumber \\
    ={}&{}
    {d_{\alpha\alpha}}\alpha\alpha(\k_1,\k_2,\k_3) 
	-\frac{4}{7}{d_{\alpha\gamma}}\alpha\gamma(\k_1,\k_2,\k_3) 
	-\frac{2}{21}\left(
        {d_{\gamma\gamma}}
        -6\lambda\beta_{K^2}
        \right)
    \gamma\gamma(\k_1,\k_2,\k_3) 
    \nonumber \\
    &
    +\frac19\left(   {d_{\xi}}+\frac92\beta_{K^3}   \right)
        \xi(\k_1,\k_2,\k_3)
    \nonumber\\
    &
    +{d_{\alpha\alpha-}}\alpha\alpha_-(\k_1,\k_2,\k_3) 
		+{d_{\alpha\gamma-}}\alpha\gamma_-(\k_1,\k_2,\k_3) 
		+{d_\zeta}\zeta(\k_1,\k_2,\k_3)
    \nonumber \\
    &
    +\left(
    \frac{\beta_3}{3!}+\frac29\beta_{K^3}+\frac23\beta_{\delta K^2}
   \right)
    +\left(\frac43\kappa\beta_{K^2}+\kappa\beta_2\right)
    \alpha_s(\k_2,\k_3)
    \nonumber \\
    &
    -\left(\frac{2}{7}\lambda\beta_2+\frac{8}{21}\lambda \beta_{K^2} +\frac12\beta_{K^3}+\beta_{\delta K^2}
    \right)
    \gamma(\k_2,\k_3)
    -2\kappa\beta_{K^2}\gamma\alpha(\k_1,\k_2,\k_3)
,\label{reduced Z_3}
\end{align}

Once the smoothing scale $R$ is determined, we can express the skewness and kurtosis parameters as a function of the model and bias parameters.
Given the linear power spectrum $P_L$\footnote{
We used the public code CLASS
\cite{Blas:2011rf,Lesgourgues:2011re} to derive the linear power spectrum.
}, the integrations in Eqs.~\eqref{skewness-specific} and \eqref{kurtosis-specific} can be performed to give the skewness and kurtosis parameters, which are schematically written as
\begin{align}
    S^{(a)}_g &= b_1^{-1}\left[\beta_2H^{(a)}_1 +\beta_{K^2}H^{(a)}_2
    +G^{(a)}\right]\qquad(a=0,1,2)\label{skewness-bias-ex}
    \,, \\
    K^{(a)}_g &= b_1^{-2}\left[\beta_3I^{(a)}_1
    +\beta_{K^3}I^{(a)}_2 +
    \beta_{\delta K^2}I^{(a)}_3+ J^{(a)}\right]
    \qquad
    (a=0,1,2_{\rI},2_{\rII},3).\label{kurtosis-bias-ex}
\end{align}
Here, $H^{(a)}_1$, $H^{(a)}_2$, $I^{(a)}_1$, $I^{(a)}_2$, and
$I^{(a)}_3$ are constants, which depend only on the linear power
spectrum. $G^{(a)}$ is the linear function of $\{\kappa$,
$\lambda\}$. $J^{(a)}$ is the sum of the quadratic functions of
$\{\beta_2$, $\beta_{K^2}$, $\kappa$, $\lambda\}$ and the linear
function of $\{\daa,\ldots, \dzeta\}$.
We note that in our analysis we apply the lowest-order (tree-level) approximation. To provide an accurate prediction, the higher-order terms should be taken into account. If one tries to calculate the skewness and kurtosis with higher order contributions, they would include more model and bias parameters. In this paper, we simply neglect these contributions.

Now we explicitly give the skewness and kurtosis parameters for the smoothing scale of $R=10\,\mathrm{Mpc}/h$. Coefficients for model and bias parameters in the expressions of the skewness and kurtosis are presented in Tables \ref{table:skewness} and \ref{table:kurtosis}.
For example, by using the value of the coefficients given in Table~\ref{table:skewness}, the first skewness parameter can be written as $S^{(0)}_g = b_1^{-1}\left(3.488\beta_2 +0.28\beta_{K^2} +4.814\kappa -1.249\lambda\right)$. We can obtain the other skewness and kurtosis parameters in the similar way.

\begin{table}[h]
\caption{Coefficients for the model and bias parameters in $b_1 S^{(a)}_g ~(a=0,1,2)$
}
\label{table:skewness}
\centering
\begin{tabular}{cccc}
    \hline
    & $b_1S^{(0)}_g$ & $b_1S^{(1)}_g$ & $b_1S^{(2)}_g$ \\
    \hline \hline
    $\beta_2$ &    3.488& 3.405& 4.255\\
    $\beta_{K^2}$ & 0.28& 0.2308& -0.4642\\
    $\kappa$  &    4.814&  4.862& 5.411\\
    $\lambda$ &   -1.249& -1.231&  -1.754\\
    \hline
\end{tabular}
\end{table}

\begin{table}[h]
\caption{Coefficients for the model and bias parameters in $b_1^2K^{(a)}_g ~(a=0,1,2_{\rm I}, 2_{\rm II},3)$
}
\label{table:kurtosis}
\centering
\begin{tabular}{l|rrrrr}
\hline
&$b_1^2K^{(0)}_g$&$b_1^2K^{(1)}_g$&$b_1^2K^{(2_\rI)}_{g}$& $b_1^2K^{(2_\rII)}_{g}$&$b_1^2K^{(3)}_{g}$\\
\hline \hline
$\beta_3$           &  5.951&    5.721&  7.849 &  5.729& 9.863\\
$\beta_{K^3}$       & -0.112&-0.09293&-0.1336  &-0.2991& 2.018\\
$\beta_{\delta K^2}$& 0.8749&  0.7784&  0.104 & -1.580& -5.298\\
$\kappa\beta_2$     &  71.29&   70.17&  96.51 & 62.84 &105.6\\
$\lambda\beta_2$    & -18.35&  -17.79& -25.93 &-17.50&-33.9 \\
$\kappa\beta_{K^2}$ & 7.972&7.042&3.653  &1.233&-5.139 \\
$\lambda\beta_{K^2}$& 6.764& 6.708& 10.71&8.116&18.87 \\
$\beta_2^2$         &  16.68&   16.08&  22.76 & 14.23&26.33 \\
$\beta_2\beta_{K^2}$&  3.181&   2.716&  1.231 & 1.177&-3.705 \\
$\beta_{K^2}^2$     & 0.2856&  0.2225&-0.08762&-0.6498&0.5774 \\
$\kappa^2$          &  31.05&   31.48&    45.8& 21.35&39.2 \\
$\kappa\lambda$     & -15.92&  -15.88&  -23.62&-13.41 &-25.62 \\
$\lambda^2$         &  2.098&   2.056&   3.161&1.883 &4.271 \\
$\daa$         &  14.07&   14.24&   15.66& 17.86&21.03 \\
$\daam$         &-4.332 &-4.082 &-6.899 &-0.8256&-5.918 \\
$\dag$         &  -7.58&  -7.546&  -9.718& -8.518&-13.39 \\
$\dagm$        &-2.414 &-2.262 &-3.852  &-1.933&-4.68 \\
$\dgg$         & -1.302&  -1.269&  -1.872& -1.489&-2.855 \\
$\dxi$         & 0.7595&  0.7404&   1.121& 0.9578&2.498 \\
$\dzeta$       & -1.108& -0.7863&  -4.391& 4.057&-3.648 \\
\hline
\end{tabular}
\end{table}

\subsection{Consistency relations
\label{sec:consistency_relation}}

\begin{figure}[t]
\centering
\includegraphics[width=\linewidth]{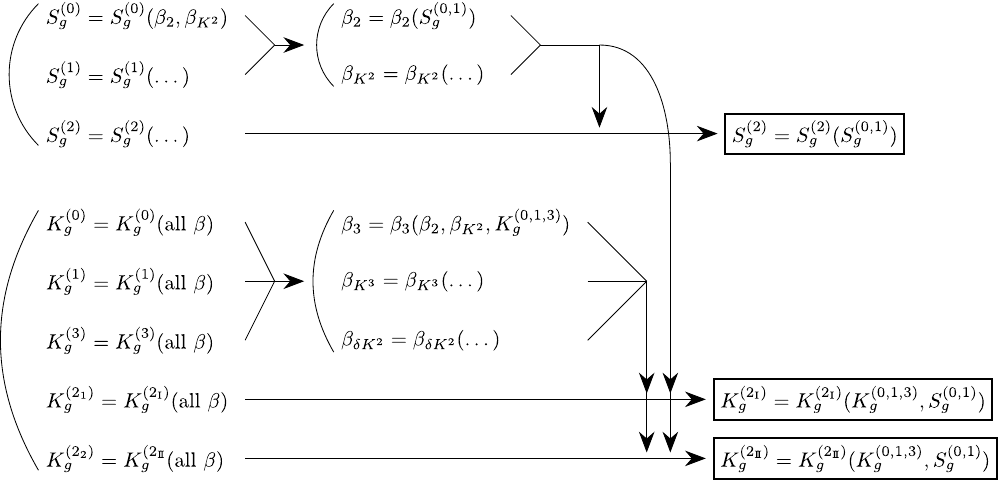}
\caption{
Sketch of
the construction of the consistency relations for the skewness and kurtosis parameters. ``All $\beta$'', $K^{(0,1,3)}_g$ and $S^{(0,1)}_g$ represent
  the following parameters respectively:
$(\beta_2,\beta_{K^2},\beta_3,\beta_{K^3},\beta_{\delta K^2})$ for ``All $\beta$", $(K^{(0)}_g,K^{(1)}_g,K^{(3)}_g)$ for $K^{(0,1,3)}_g$, and
  $(S^{0}_g,S^{(1)}_g)$ for $S^{(0,1)}_g$.  Also, $(\dots)$ indicates
  the same arguments with the above quantity.  Skewness consistency
  relation can be derived by 
  removing $\beta_2$ and $\beta_{K^2}$
  using two conditions of skewness parameters. 
  On the other hand, the kurtosis
  consistency relations can be derived by 
  removing all bias
  parameters using three conditions of kurtosis parameters and two
  skewness parameters.}
\label{fig:consistency}
\end{figure}

In the previous section, the expressions for the skewness and kurtosis equations are obtained, from which we can derive the consistency relations for the skewness and kurtosis parameters. An important observation from Eqs.~\eqref{skewness-bias-ex} and \eqref{kurtosis-bias-ex} is that several bias parameters linearly appear in the skewness and kurtosis equations. Hence, such bias parameters can be solved and written in terms of other quantities. We then eliminate the biases by substituting the obtained equations into the remaining skewness and kurtosis equations to construct the consistency relations including only the observables (the skewness and the kurtosis) and model parameters.
Fig.~\ref{fig:consistency} illustrates how we can obtain the consistency relations. We first solve Eqs.~\eqref{skewness-bias-ex} and \eqref{kurtosis-bias-ex} and rewrite the bias parameters in terms of two skewness ($S^{(0)}_g$ and $S^{(1)}_g$) and three kurtosis parameters ($K^{(0)}_g$, $K^{(1)}_g$ and $K^{(3)}_g$), which are schematically written as
\begin{align}
&\beta_i = \beta_i(S^{(0)}_g, S^{(1)}_g, \kappa, \lambda) 
\qquad (i=2,K^2) \,,
\label{skew-bias-condition} \\
& \beta_j= 
    \beta_j(K^{(0)}_g, K^{(1)}_g, K^{(3)}_g,\beta_2,\beta_{K^2},\kappa,\lambda, \daa,\ldots ,\dzeta)
    \quad (j=3, {K^3},{\delta K^2}) \,.
    \label{kurt-bias-condition}
\end{align}  
We then substitute Eq.~\eqref{skew-bias-condition} into the remaining skewness relation $S^{(2)}_g=S^{(2)}_g(\beta_2, \beta_{K^2}, \kappa, \lambda)$ to obtain the consistency relation among the skewness parameters:
\begin{align}
    S^{(2)}_g=S^{(2)}_g(S^{(0)}_g, S^{(1)}_g, \kappa, \lambda)  \,,
    \label{general-skew-consistency}
\end{align}
whose explicit form will be given later. 

We can also obtain the two consistency relations for the kurtosis
parameters by inserting Eq.~\eqref{kurt-bias-condition} into $
K^{(a)}_g=K^{(a)}_g(\beta_2,\beta_K^2,\beta_3,\beta_{K^3},\beta_{\delta K^2},\daa\ldots \dzeta) $ with $a=2_{\rI},2_{\rII}$, from which one can find:
\begin{align}
K^{(a)}_g=K^{(a)}_g(\beta_2,\beta_K^2,K^{(0)}_g,K^{(1)}_g,K^{(3)}_g,\daa\ldots \dzeta)
=
K^{(a)}_g(S^{(0)}_g,S^{(1)}_g,K^{(0)}_g,K^{(1)}_g,K^{(3)}_g,\daa\ldots \dzeta).
\end{align}
Therefore, we can derive three consistency relations: one for the skewness parameters 
and two for the kurtosis parameters.

\begin{figure}[t]
\centering
\includegraphics[width=\linewidth]{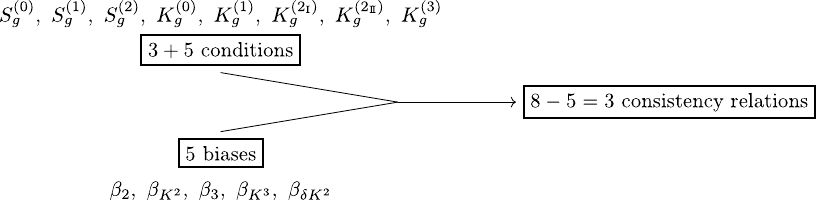}
\label{fig:solv_consistency}
\caption{Using the expressions of skewness and kurtosis parameters,
  Eqs.\eqref{skewness-bias-ex} and \eqref{kurtosis-bias-ex}, we can
  eliminate five unknown bias parameters and derive three consistency
  relations.}
\end{figure}

Now let us demonstrate how we obtain
explicit expressions for the bias parameters, and then derive the consistency relations with 
the smoothing scale $R=10~\mathrm{Mpc}/h$. 
Since the bias parameters can be given as a function of the skewness, kurtosis and model parameters, we give the coefficients of these parameters for
the second-order and third-order bias in 
Tables~\ref{table:bias-second-order} and \ref{table:bias-third-order} 
respectively. For example, by using the coefficients given in Table~\ref{table:bias-second-order}, the second-order bias parameters $\beta_2$ and $\beta_{K^2}$ are written as 
\begin{align}
    &\beta_2 =-1.687 \kappa +0.381 \lambda -1.555 b_ 1 S^{(0)}_g+1.886 b_ 1 S^{(1)}_g  \,, \\
    &\beta_{K^2} =3.822 \kappa -0.2857 \lambda +22.93 b_1 S^{(0)}_g-23.49 b_1 S^{(1)}_g \,.
\end{align}
Similar expressions can also be obtained for other bias parameters.

\begin{table}[t]
\caption{Coefficients for $\{\kappa,\lambda,b_1S^{(0)}_g,b_1S^{(1)}_g\}$ appearing in the expressions of
$\beta_2$ and $\beta_{K^2}$
with $R=10~\mathrm{Mpc}/h $.}
\label{table:bias-second-order}
\centering
\begin{tabular}{ccc}
    \hline
    & $\beta_2$ & $\beta_{K^2}$ \\
    \hline \hline
    $\kappa$    & -1.687    & 3.822\\
    $\lambda$   & 0.381     & -0.2857\\
    $b_1S^{(0)}_g$ &-1.555  &  22.93\\
    $b_1S^{(1)}_g$ &1.886   & -23.49\\
    \hline
\end{tabular}
\end{table}
\begin{table}[t]
\caption{
Coefficients for $\{\kappa^2,\kappa\lambda,\lambda^2,\daa \ldots
  \dzeta,b_1\kappa S_g^{(0)} \ldots b_1^2K^{(0,1,3)}_g\}$ appearing in the expressions of
$\beta_3,~\beta_{K^3}$ and $\beta_{\delta K^2}$
with $R=10~\mathrm{Mpc}/h $.
}
\label{table:bias-third-order}
\centering
\begin{tabular}{llll}
\hline
& $\beta_3$ & $\beta_{K^3}$ & $\beta_{\delta K^2}$ \\
\hline \hline
$\kappa^2$             &14.55  & -335.3 &-110.3\\
$\kappa\lambda$        &-5.402  &82.23  &28.43\\
$\lambda^2$            &0.4985 &-3.337  &-1.267\\
$\daa$                 &6.051 &128.0    &41.46\\
$\daam$                &0.3363 &12.48   &4.261 \\
$\dag$                 &2.577  &-44.02  &-14.5\\
$\dagm$                &0.08687 &11.43  &3.631 \\
$\dgg$                 &0.2908 &-1.947  &-0.7391\\ 
$\dxi$                 &-0.1481 &-0.2222&0.1111\\
$\dzeta$               &-1.291 &51.96   &16.70\\
$b_1\kappa S^{(0)}_g$  &42.56  &-1855.  &-604.4\\
$b_1\kappa S^{(1)}_g$  &-45.57   &1931. &-627.3\\
$b_1\lambda S^{(0)}_g$ &-6.847 & 267.9   &93.42\\
$b_1\lambda S^{(1)}_g$ &7.349 &-274.4    &-95.78\\
$b_1^2{S^{(0)}_g}^2$        &73.98   &-3256.  &-1008.\\
$b_1^2{S^{(0)}_gS^{(1)}_g}$ &-142.2   &6357.   &1954.\\
$b_1^2{S^{(1)}_g}^2$        &67.73   &-3094  &-943.8\\
$b_1^2{K^{(0)}_g}$          &-4.713   &166.0  &54.46\\
$b_1^2{K^{(1)}_g}$          &5.122 &-175.5   &-57.31\\
$b_1^2{K^{(3)}_g}$          &-0.02576&1.629  &0.3838\\
\hline
\end{tabular}
\end{table}

By plugging the expressions for these bias parameters into the skewness and kurtosis relations, $S^{(2)}_g$\,, $K^{(2_{\rI})}_g$\,, $K^{(2_{\rII})}_g$, 
we finally obtain three consistency relations.
The explicit forms of these are given by, 
for the smoothing scale of $R=10~{\rm Mpc}/h$, 
\begin{align}
	&0=-17.26S^{(0)}_g+18.93S^{(1)}_g-\frac{3.542 \kappa }{{b_1}}-{S^{(2)}_g}
        \label{skewness-consistency}
	\\
	&0=b_1^2\Bigl(
        53.52 {K^{(0)}_g}
        -57.69 {K^{(1)}_g}
        +{K^{(2_\rI)}_{g}}
        +0.380 {K^{(3)}_g}
	\nonumber \\
        &\qquad\qquad
        -876.0 {S^{(0)}_g}^2 
        -1703. {S^{(1)}_g} {S^{(0)}_g}
        -825.0 {S^{(1)}_g}^2
        \Bigr) \nonumber \\
	&\quad
	+b_1\left(
            -502.0 {S^{(0)}_g} \kappa 
            +525.7 {S^{(1)}_g} \kappa 
            +81.76 {S^{(0)}_g} \lambda 
            -83.75 {S^{(1)}_g} \lambda
        \right)         \nonumber \\
	&\quad
            -100.0 \kappa ^2
            +25.88 \lambda  \kappa 
            -1.018 \lambda ^2
	\nonumber \\
        &\quad
            +44.63 \daa 
            +5.484\daam
            -14.88 \dag
            +4.320 \dagm
            +0.5941 {\dgg}
            +19.73 {\dzeta} 
        \label{kurtosis-consistency-1}
        \\
	&0=b_1^2\Bigl(
            +162.7 {K^{(0)}_g}
            -172.4 {K^{(1)}_g}
            +{K^{(2_\rII)}_{g}}
            +1.241 {K^{(3)}_g}      
        \nonumber \\
        &\qquad \qquad
            -2641. {S^{(0)}_g}^2
            +5093. {S^{(1)}_g} {S^{(0)}_g}
            -2444. {S^{(1)}_g}^2
        \Bigr)\nonumber \\
	&\quad
	+b_1\left(
            -1592. {S^{(0)}_g} \kappa 
            +1659. {S^{(1)}_g} \kappa 
            +265.2 {S^{(0)}_g} \lambda 
            -271.6 {S^{(1)}_g} \lambda 
        \right)\nonumber \\
	&\quad
    	-301.4 \kappa ^2
            +79.99 \lambda  \kappa 
            -3.303\lambda ^2
        \nonumber \\
        &\quad
            +120.6 {\daa}
            +9.363\daam
            -42.32 {\dag}
            +10.59\dagm
            -1.927 {\dgg}
            +45.26 {\dzeta}
        \label{kurtosis-consistency-2}
\end{align}
We note that the first one of the above expressions 
has been already derived in \cite{Yamauchi:2022fss}.
When the smoothing scale is specified, the skewness and kurtosis parameters appearing in the consistency relations \eqref{skewness-consistency}--\eqref{kurtosis-consistency-2} can be directly evaluated from observational data of galaxy distributions by using the definition \eqref{skewness-kurtosis-def}.
It should be noticed that the bias parameters that are difficult to determine from observations, namely
$\beta_2,\beta_{K^2},\beta_3,\beta_{K^3},\textrm{ and
}\beta_{\delta K^2}$, can be safely removed, and then the consistency relations only include $b_1$, model parameters, skewness, and kurtosis.  
Therefore, these consistency relations can be used to constrain the model parameters characterizing the underlying gravity theory such as the DHOST theory without suffering from the nonlinear biases.

\begin{figure}[t]
\centering
\includegraphics[width=0.99\linewidth]{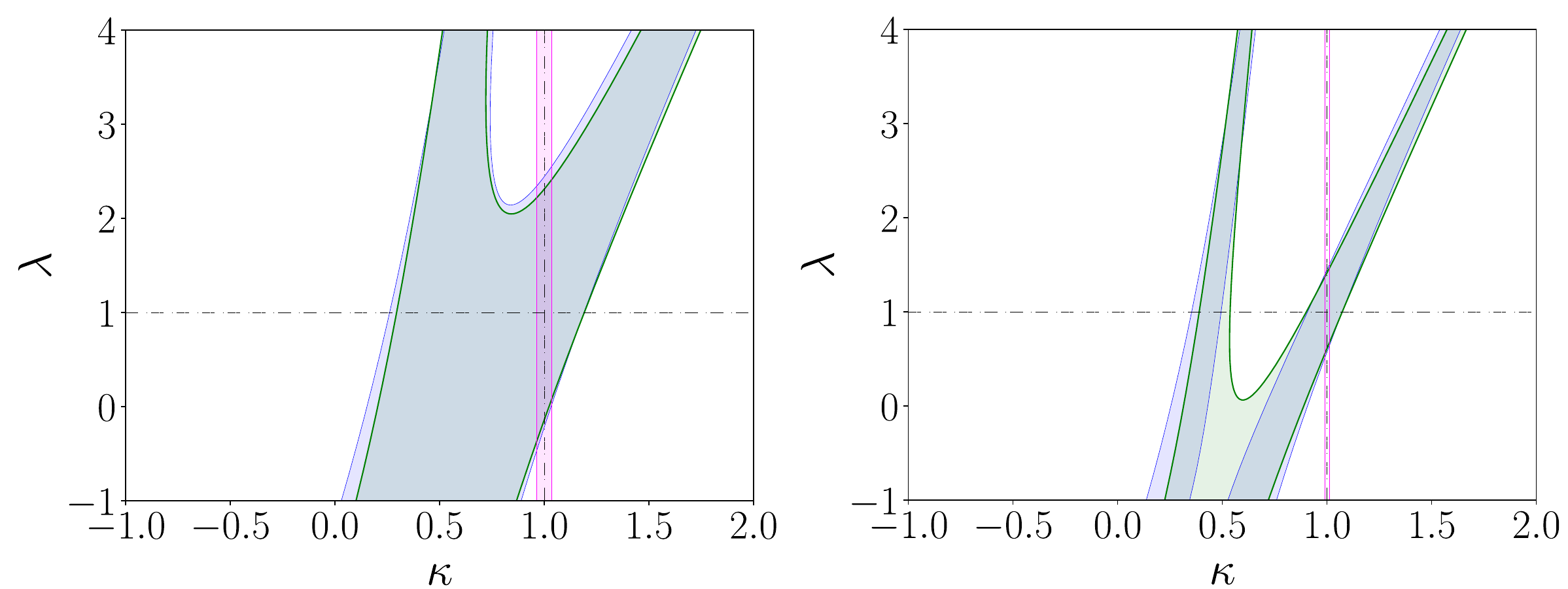}
\caption{Forecast of the constraint on the model parameters $\kappa$ and $\lambda$ by using the skewness and the kurtosis consistency relations Eqs.~\eqref{skewness-consistency},
  \eqref{kurtosis-consistency-1} and \eqref{kurtosis-consistency-2}.
  In these plots, other model parameters
  $\daa,\dag,\dgg,\dxi \textrm{and } \dzeta$ are assumed to be the EdS's values.
  The skewness and kurtosis parameters are taken to be the EdS values (see \cite{Matsubara:2020knr})
with the 1-$\sigma$ (left plot) and $1/3$-$\sigma$ (right plot) errors.
  $\kappa=\lambda=1$ (black dashed line crossing point) corresponds to the ones in the EdS model.}
  \label{fig:constraint}
\end{figure}

Before closing this section, let us show the potential impact of the use of the consistency relations to constrain the gravity theories. We now consider the following functions:
\begin{align}
    \mathcal S = \textrm{RHS of \eqref{skewness-consistency}},\\
    \mathcal K_1 = \textrm{RHS of \eqref{kurtosis-consistency-1}},\\
    \mathcal K_2 = \textrm{RHS of \eqref{kurtosis-consistency-2}}.
\end{align}
We show in Fig.~\ref{fig:constraint}
the allowed parameter regions obtained from the derived consistency relations with the expected errors of the skewness and kurtosis parameters. The consistency relations for the skewness ${\cal S}$ (magenta) and kurtosis ${\cal K}_1,{\cal K}_2$  (blue and green) with the 1 (left) and 1/3 (right) $\sigma$ errors of each variable estimated in N-body simulations (1 $h^{-3}$Gpc$^3$ $\times$ 300 realizations) conducted in Ref.~\cite{Matsubara:2020knr} are plotted.
We take the values in the EdS as the fiducial parameters.
We found that the overlap region, where all these constraints are satisfied, becomes much smaller than the individual allowed region.
In particular, in the extreme settings, 
one can exclude $\lambda =0$ by the combined analysis for the skewness and kurtosis consistency relations in future observational data, as shown in the right plot of Fig.~\ref{fig:constraint}.
The allowed overlapped region indicates the constraints on $\kappa$ and $\lambda$ as
$0.92\lesssim\kappa\lesssim 1.08$ and
$0.62 \lesssim \lambda \lesssim 1.41$.
Our results show that the combined analysis of the skewness and kurtosis parameters and the consistency relations will be able to provide tighter constraints on the theory of gravity. 

\section{Conclusion  \label{sec:conclusion}}

We derived the consistency relations of the skewness and kurtosis parameters, based on the perturbation theory of the matter density fluctuations, in which the information of the underlying gravity theories is encoded. 
We first developed the formula of the skewness and kurtosis parameters with several spatial derivatives based on the third-order matter perturbation theory.
The presence of the nonlinear galaxy bias leads to observational uncertainties.
Therefore, it would be an important task for future observations to construct consistency relations among the observables without suffering from nonlinear galaxy biases.
We found that several nonlinear bias parameters linearly appear in the skewness and kurtosis parameters, hence such biases can be rewritten in terms of the observables and safely eliminated. 
In this paper, based on the resultant expressions, we reproduced one consistency for the skewness parameters and two new consistency relations
for the kurtosis parameters. 
These three consistency relations are applicable to 
a very wide class of gravity theories such as the DHOST theory.
It should be noted that, 
by construction, these consistency relations do not include the nonlinear galaxy biases.

With the formulas derived in this paper,
we demonstrated the potential impact of the combined analysis of the consistency relations for the skewness and kurtosis
to constrain the gravity theories.
When we consider the future observations with the extremely large survey volume
and the smoothing scale $R=10\,{\rm Mpc}/h$, we found
the constraints on $\kappa$ and $\lambda$ as
$0.92\lesssim\kappa\lesssim 1.08$ and
$0.62 \lesssim \lambda \lesssim 1.41$ 
by considering the allowed regions obtained by the combined analysis with three consistency relations where third-order model parameters are fixed by EdS gravity and 
the errors of the skewness and kurtosis are estimated by the N-body simulations.

As a final remark, we comment on the possible impacts of some of our assumptions in the analysis of this paper. 
First, we assumed the local bias model with constant coefficients and a negligible velocity term. Second, we gave a
smoothing prescription after applying bias models.  
These may affect the accuracy of model constraints.
Third, when we calculated the skewness and kurtosis parameters, we assumed the lowest-order (tree-level) approximation. If we need more accurate predictions, we may have to take account of the higher-order contributions for the matter perturbations.  
Finally, we focused only on the analysis in 
real space, but the actual galaxy number fluctuations should be defined in redshift space. It would be interesting to extend our analysis to the redshift space (see \cite{Yamauchi:2022fss} for the skewness consistency), which are left for a future issue.

\section*{Acknowledgements}
This work was supported by JSPS KAKENHI Grant Number 19K03874~(TT),
23K17691~(TT), 22K03627~(DY), 23K25868~(DY) and MEXT KAKENHI
23H04515~(TT),
JP19K03835 (TM) and
21H03403 (TM).

\appendix

\pagebreak
\bibliography{refs}

\end{document}